\documentstyle[preprint,eqsecnum,aps]{revtex}

\begin{document}
\draft
\preprint{HEP/123-qed}

\title{Relativistic predictions of quasielastic proton-nucleus
spin observables based on a complete Lorentz invariant representation
of the NN scattering matrix}

\author{B.I.S. van der Ventel, G.C. Hillhouse, and P.R. De Kock}
\address{Physics Department, University of Stellenbosch, 
Stellenbosch, 7600, South Africa}
\date{\today}
\maketitle

\begin{abstract}
Within the framework of the relativistic plane wave impulse 
approximation (RPWIA), complete sets of quasielastic $(\vec{p}, \vec{p}^{\; '})$
and $(\vec{p}, \vec{n})$ spin observables are calculated employing 
a general Lorentz invariant representation of the NN scattering matrix
(referred to as the IA2 representation). The use of a complete 
representation eliminates the arbitrariness of a previously-used 
five-term parameterization (commonly called the IA1 representation) 
and allows for the correct incorporation of 
effective-mass-type medium effects within the RPWIA framework and 
within the context of the Walecka model. For quasielastic scattering
from a $^{40}$Ca target at incident proton energies between 200 and
500 MeV, we investigate the sensitivity of complete sets of spin observables to 
effective nucleon masses for both IA1 and IA2 representations. 
In general it is seen that the IA1 representation may overestimate the importance of nuclear medium effects, whereas the
IA2-based predictions nearly correspond to values for
free nucleon-nucleon scattering.
\end{abstract}

\narrowtext

\section{Introduction}
\label{section_introduction}
In a previous paper we developed a relativistic plane wave 
model for studying medium modifications of the nucleon-nucleon (NN)
interaction via complete sets of spin observables 
for quasielastic $(\vec{p}, \vec{p}^{\; '}) $ and $(\vec{p}, \vec{n})$
scattering \cite{vanderVentel_PRC60_99}. A systematic survey of 
the predictive  power of the latter model compared to 
experimental data will be presented in this paper.

The main aspect of our model is the use of a general Lorentz invariant representation of the NN scattering matrix referred to as the IA2 representation. This complete expansion of the interaction matrix contains 44 independent invariant amplitudes consistent with parity and time-reversal invariance as well as charge symmetry together with the on-mass-shell condition for the external nucleons \cite{Tjon_PRC32_85,Tjon_PRC35_87}. Five of the 44 amplitudes are determined from free NN scattering data and are therefore identical to the amplitudes employed in the previously-used five-term parameterization of the NN scattering matrix referred to as the IA1 representation. The remaining 39 amplitudes may be obtained via solution of the Bethe-Salpeter equation employing a one-boson exchange model (with pseudovector pion-nucleon coupling) for the NN interaction \cite{Tjon_PRC32_85,Tjon_PRC35_87,vanFaassen_PRC28_83,vanFaassen_PRC30_84}. The use of a complete set of NN amplitudes eliminates ambiguities inherent in the IA1 representation.
The effect of the nuclear medium on the scattering wave functions is incorporated by replacing free nucleon masses in the Dirac spinors 
with smaller effective projectile and target nucleon masses within the 
context of the relativistic mean field approximation of 
Serot and Walecka \cite{Se86}. Experimental data on quasielastic
spin observables suggest that nuclear shell effects are
unimportant, and hence the target nucleus is treated as a non-interacting Fermi gas. 

One of the great triumphs of Dirac phenomenology has been 
the successful prediction of the analyzing power for quasielastic
$^{40}\mbox{Ca}(\vec{p}, \vec{p}^{\; '} ) $ scattering 
at 500 MeV based on the IA1 representation of the NN
interaction within the framework of
a simple relativistic plane wave model \cite{Ho88}. 
The latter success is achieved
by replacing free nucleon masses with
effective nucleon masses in the Dirac spinors, thus enhancing the
lower components of the Dirac spinors and resulting in a reduction of
the analyzing power relative to the value for free scattering:
this reduction has been called
a "relativistic signature" since no mechanism has been found
for its explanation within the framework of the conventional
nonrelativistic Schr\"{o}dinger equation.
Despite the successful prediction of the analyzing power, 
however, the relativistic IA1-based model yields 
inconsistencies in the sense that quasielastic 
$ ( \vec{p}, \vec{p}^{\; '} ) $ and $ ( \vec{p}, \vec{n} ) $ 
spin observables prefer different five-term representations 
of the NN scattering matrix. As already explained, a more rigorous 
and unambiguous approach must be based on the IA2 representation 
of the scattering matrix within the relativistic plane wave impulse approximation. In Ref. \cite{vanderVentel_PRC60_99} we showed 
that the inclusion of effective
masses within the IA2 representation fails to reproduce the large
quenching effect predicted by the IA1 representation for the
$^{40}\mbox{Ca}(\vec{p}, \vec{p}^{\; '} ) $ analyzing power at 500 MeV. Hence,
we concluded that any large deviations of 
spin observables relative to the corresponding free values 
are merely artifacts of using an incorrect IA1 representation 
of the NN scattering matrix, and consequently other effects, 
such as distortions and multiple scattering, should be 
considered as possible candidates for reproducing the 500 MeV 
analyzing power within the IA2 representation. 

The question now arises as to how IA2-based
predictions compare to data at energies lower than 500 MeV for a
range of scattering angles, and how do they compare to the 
corresponding IA1-based predictions. In principle all calculations should be based on the more rigorous IA2 representation, however, for comparison to previous predictions, the IA1-based calculations are included. In addressing the above questions, we attempt to fully understand the role of 
effective-mass-type medium 
effects on spin observables before attempting to
incorporate additional effects into our relativistic model.
The aim of this paper, therefore, is to perform a systematic study 
of the predictive power of IA2-based model compared to the published 
quasielastic polarization data listed in Table \ref{studied_reactions}.
The following questions will also be addressed:
\begin{itemize}
\item
How successful is the effective mass concept, inherent to
Dirac phenomenology, in describing quasielastic 
$ (\vec{p}, \vec{p}^{\; '} ) $ and $ (\vec{p}, \vec{n} ) $ 
scattering data?
\item
How do numerical results based on the IA2 representation of 
the NN scattering matrix compare to those utilizing the incomplete (and therefore ambiguous) IA1 representation?
\end{itemize}
In Sec. \ref{section_IA1_versus_IA2_at_QEP} the sensitivity
of complete sets of quasielastic $ (\vec{p}, \vec{p}^{\; '} ) $ 
and $ (\vec{p}, \vec{n} ) $ spin observables is investigated
with respect to a range of different effective projectile and target 
nucleon masses for both IA1- and IA2-based models. In addition,
calculations based on optimal combinations of effective projectile
and target nucleon masses are also compared to spin observable data
at the centroid of the quasielastic peak. Our main conclusions
are presented in Sec. \ref{section_conclusions}.

\section{Sensitivity of spin observables to effective masses}
\label{section_IA1_versus_IA2_at_QEP}
In Ref. \cite{vanderVentel_PRC60_99} it was 
shown that an IA2-based prediction fails to
reproduce the $^{40}$Ca$ ( \vec{p}, \vec{p}^{\; '} ) $ analyzing
power at an incident energy of 500 MeV. In order to 
give an initial feeling for the predictive power 
of our model, the latter reference employed values 
of the effective nucleon masses which were theoretically 
extracted by Hillhouse and De Kock \cite{hillhouse_PRC49_94}. 
However, the question arises as to whether 
other combinations of physically acceptable
effective  projectile and target nucleon masses exist, which 
provide a better description of the analyzing power. Furthermore, one
can also ask whether the latter combination still provides a good 
description of all the other spin observables, and if not, whether one can find a combination of physically acceptable effective masses which 
reproduce a complete set of spin observables.

Table \ref{studied_reactions} lists all the reactions for 
which calculations are done. In this paper we only present
the results for the $^{40}$Ca target since this is 
representative of the results which were obtained for 
all the other target nuclei. Results for the 
last four reactions can be found in Ref. \cite{vanderVentel_PhD_99}. 
Complete sets of spin observable data exist for all the 
energies and targets used, except $^{40}$Ca$ ( \vec{p}, \vec{n} ) $ 
at $ T_{lab} \; = \; 495 $ {\rm MeV} for which no analyzing
power data are available and $^{40}$Ca $ ( \vec{p}, \vec{p}^{\; '} ) $ 
at $ T_{lab} \; = \; 200 $ {\rm MeV} for which only $ A_{y} $ and 
$ D_{nn} $ data are available. The reaction 
$^{40}$Ca$ ( \vec{p}, \vec{n} ) $ at 
$ T_{lab} \; = \; 495 $ {\rm MeV} is included since data exist 
at two different laboratory scattering angles and furthermore it is 
complementary to the reaction $^{40}$Ca$ (\vec{p}, \vec{p}^{\; '} ) $ 
at $ T_{lab} \; = \; 500 $ {\rm MeV}. The $ (\vec{p}, \vec{p}^{\; '} ) $ 
data at $ T_{lab} \; = \; 200 $ {\rm MeV} are complementary to the 
$ ( \vec{p}, \vec{n} ) $ data at $ T_{lab} \; = \; 200 $ {\rm MeV} 
and are therefore also included.

\subsection{Effective mass bands}
To answer the above questions, we introduce the concept of an 
effective mass band in this section, which serves to demonstrate 
the sensitivity of spin observables to different combinations 
of effective masses for projectile and target nucleons for both 
IA1- and IA2-based models. 
In principle the effective masses can be calculated 
theoretically following a procedure similar to that 
outlined in Ref. \cite{hillhouse_PRC49_94}, however, 
the effective masses are now considered as free parameters which are
varied, in step sizes of 0.01, over the following range of 
physically acceptable values: 
\begin{eqnarray}
\label{effecmass_range}
 ( 0.50; 0.50 ) \, \leq & 
 \displaystyle{( \frac{M_{1}}{M}; \frac{M_{2}}{M} )} \, 
\leq & (1.0; 1.0).
\end{eqnarray}
$ M $ denotes the free nucleon mass, and $ M_{1} $ and $ M_{2} $ the effective
projectile and target nucleon masses respectively. The lower limit of 0.50 corresponds
to the effective nucleon mass in infinite nuclear matter \cite{Se86}.
For the purpose of this exercise we focus on values of
the spin observables at an excitation energy corresponding
to the centroid of the quasielastic peak in the unpolarized
inclusive excitation spectrum. For different laboratory 
scattering angles empirical data for quasielastic spin 
observables are relatively constant as a function of nuclear 
excitation energy at the momentum transfers of interest 
($|\vec{q}\,|\, >\, 0.5$ fm $^{-1}$). Hence
the trends displayed by observables at the quasielastic peak
will be representative of the behavior of spin observables as a
function of the energy transferred to the nucleus.

We now introduce the concept of an effective mass band for a 
particular reaction at a fixed incident energy as a function
of laboratory scattering angle. Let 
$D_{i' j} ( \omega, \theta_{lab}, \frac{M_{1}}{M}, \frac{M_{2}}{M})$ 
denote a particular spin observable from the complete set 
$ \{ A_{y}, D_{\ell' \ell}, D_{s' s}, D_{\ell' s}, D_{s' \ell}, D_{nn} \} $ 
with $ D_{0n} \, \equiv \, A_{y} $, where $ \omega $ is the 
energy transferred to the nucleus and $ \theta_{lab} $ is the 
laboratory scattering angle. For the IA2-based model, the procedure 
for calculating quasielastic spin observables is outlined
in Ref. \cite{vanderVentel_PRC60_99}. For the IA1 representation 
of the NN scattering matrix we employ the phenomenological 
Horowitz-Love-Franey \cite{horowitz_PRC31_85} model with 
pseudovector pion-nucleon coupling as explained in Ref. \cite{hillhouse_PRC49_94}. In order to do the IA1 calculations, 
new Horowitz-Love-Franey parameters were generated for the 
energy range of 80 to 195 MeV in steps of 5 MeV
\cite{hillhouse_PhD_99}, and for laboratory energies 
higher than 200 MeV we employed the Maxwell parameterization of the 
NN amplitudes \cite{maxwell_NuclPhysA600_96,maxwell_NuclPhysA638_98}.

In order to generate the effective mass bands,
the spin observables are first calculated as 
a function of $ \omega $ (for fixed $ \theta_{lab} $), 
and then the value of the particular spin observable 
is extracted {\em at the quasielastic peak}, i.e.
\begin{eqnarray}
\label{spin_observable_at_peak}
 D_{i' j}^{( peak)} ( \theta_{lab},\displaystyle{\frac{M_{1}}{M}}, \displaystyle{\frac{M_{2}}{M}} ) 
 & = &
 D_{i' j} ( \omega \, = \, \omega_{peak}, \theta_{lab},\displaystyle{\frac{M_{1}}{M}}, \displaystyle{\frac{M_{2}}{M}} ).
\end{eqnarray}
where $ \omega_{peak} $ is the experimental value of the 
energy transfer associated with the centroid of the quasielastic peak.
For a {\em fixed} $ \theta_{lab} $, each spin observable is calculated successively for each of the different effective mass combinations 
in Eq. (\ref{effecmass_range}). This is repeated for 
$ 10^{\circ} \, \leq \, \theta_{lab} \, \leq \, 60^{\circ} $ and 
therefore each effective mass combination generates a curve as a 
function of $ \theta_{lab} $. Instead of plotting all the different 
curves on one graph, we calculate, for a {\em fixed} $\theta_{lab}$, 
the minimum and maximum values for a particular spin observable:
\begin{eqnarray}
\label{spinobservable_minimum}
 \left( D_{i' j}^{(peak)} \right)_{min} ( \theta_{lab} ) 
 & = &
 {\rm Min} [ D_{i' j}^{(peak)} ( \theta_{lab}, 1.0; 0.5); D_{i' j}^{(peak)} ( \theta_{lab}, 1.0; 0.6); \cdots D_{i' j}^{(peak)} ( \theta_{lab}, 1.0; 1.0) ]
\nonumber\\
\label{spinobservable_maximum}
 \left( D_{i' j}^{(peak)} \right)_{max} ( \theta_{lab} ) 
 & = &
 {\rm Max} [ D_{i' j}^{(peak)} ( \theta_{lab}, 1.0; 0.5); D_{i' j}^{(peak)} ( \theta_{lab}, 1.0; 0.6); \cdots D_{i' j}^{(peak)} ( \theta_{lab}, 1.0; 1.0) ]\,.
\end{eqnarray}
As $ \theta_{lab} $ varies between $ 10^{\circ} $ and 
$ 60^{\circ} $ $ \left( D_{i' j}^{(peak)} \right)_{min} 
( \theta_{lab} ) $ traces out a lower curve and 
$ \left( D_{i' j}^{(peak)} \right)_{max} ( \theta_{lab} ) $ 
traces out an upper curve on the graph. All effective mass 
combinations given by Eq. (\ref{effecmass_range}) lie between 
these limits, and this (as a function of scattering angle) 
forms an effective mass band for each spin observable. Effective 
mass bands for both IA1 and IA2 representations of the 
relativistic NN scattering matrix are presented in Figs. \ref{fig_ca40_500pp19_QEP} to \ref{fig_ca40_200pn_cheri_QEP} for
$ ( \vec{p}, \vec{p}^{\; '} ) $ and $ ( \vec{p}, \vec{p}^{\; '} ) $
scattering from a $^{40}$Ca nucleus at incident energies
of 200 and 500 MeV. Similar figures for the other reactions listed in 
Table \ref{studied_reactions} can be found in Ref. \cite{vanderVentel_PhD_99}.
The energy range is chosen to correspond to polarized proton energies
of interest to experimental programs at facilities such as the National Accelerator Centre (Faure, South Africa) and The Research Center for 
Nuclear Physics (Osaka, Japan).
The IA1 and IA2-based effective mass bands are denoted
by the straight-line-hatch and dotted-hatch patterns
respectively. The solid circles represent the experimental 
values extracted at the quasielastic peak for a specific 
laboratory scattering angle: the data are taken from 
references cited in Table \ref{studied_reactions}.

The effective mass bands for the different reactions in Figs. \ref{fig_ca40_500pp19_QEP} to \ref{fig_ca40_200pn_cheri_QEP} are self-explanatory: if a data point falls outside 
a band, it means that no effective mass combination can describe 
that particular point; Rather one must consider other effects
such as distortions, multiple scattering or recoil effects
in an attempt to reproduce the data. The width of a band also 
gives an indication of the expected medium effect on a 
particular spin observable; If the band is wide, then this 
spin observable is sensitive to a variation 
in effective masses and it may exhibit a large deviation from the 
free mass calculation, i.e. a large medium effect. Vice versa 
if the band is very narrow. The advantage of the effective mass 
band plots is that they immediately give an indication of whether 
a particular spin observable can be described via the concept of 
an effective-mass. 

Although Figs. \ref{fig_ca40_500pp19_QEP} to 
\ref{fig_ca40_200pn_cheri_QEP} speak for themselves, we briefly
highlight the main results. For both $ ( \vec{p}, \vec{p}^{\; '} ) $ 
and $ ( \vec{p}, \vec{n}) $ scattering the IA1 bands are 
broader than the IA2 bands, indicating the that the IA1 representation
severely overestimates the role of effective-mass-type medium effects
for quasielastic scattering. In addition, as the energy is lowered, the IA1 bands become broader for $ ( \vec{p}, \vec{p}^{\; '} ) $ 
scattering. For $ ( \vec{p}, \vec{p}^{\; '} ) $ scattering at
200 MeV (Fig. \ref{fig_ca40_200pp30_QEP}) both representations fail
to describe $A_{y}$ and $D_{n' n}$ indicating that other effects
(other than effective-mass-type effects) may play a more important role
at low incident energies. Note that for $ ( \vec{p}, \vec{p}^{\; '} )$
scattering at both 200 and 500 MeV (Figs. \ref{fig_ca40_500pp19_QEP}
and \ref{fig_ca40_200pp30_QEP}) the IA2-based model fails to
reproduce the $ A_{y} $ and $ D_{nn} $ data. Fig. \ref{fig_ca40_200pn_cheri_QEP}
for $ (\vec{p}, \vec{n}) $ scattering at 200 MeV clearly illustrates
the danger of interpreting medium effects within the IA1 representation:
the band for the ambiguous IA1 representation includes the data points for both
$D_{s' l}$ and $D_{l' s}$ spin observables, whereas the more rigorous
IA2-based band excludes these data points.

\subsection{Optimal effective mass combinations}
Next we extract that combination of effective projectile and 
target nucleon masses which best describes a complete set of 
spin observables for a range of scattering angles at a fixed 
incident energy. The systematics of these so-called 
optimal effective masses is studied for both IA1- and IA2-based models
and also compared to values calculated from empirical scalar potentials in an eikonal approximation \cite{hillhouse_PRC49_94}.

We start by defining:
\begin{eqnarray}
\label{chisqrd_def}
 \displaystyle{\Delta} ( \frac{M_{1}}{M}, \frac{M_{2}}{M} )
 & = &
 \sum_{ i = 1 }^{ n_{1} } \; \sum_{ j = 1 }^{ n_{2} } \;
 ( w_{theory}^{(j)} ( \theta_{i} ) - w_{expr}^{(j)} ( \theta_{i} ) )^{2}
\end{eqnarray}
where $ w_{theory}^{(j)} ( \theta_{i} ) $ is the theoretical value 
of the spin observable evaluated at the laboratory scattering angle 
$ \theta_{i} $ at which the experimental data are available. Similarly
$ w_{expr}^{(j)} ( \theta_{i} ) $ is the experimental value of the 
spin observable. $ n_{1} $ and $ n_{2} $ denote the number of 
laboratory scattering angles at which data exist and the number of 
spin observables which were measured, respectively. For example, 
for the reaction  $^{40}$Ca$ ( \vec{p}, \vec{p}^{\; '} ) $ at
$ T_{lab} \; = \; 500 $ {\rm MeV},  $ n_{1} \; = \; 1 $ (data 
measured only at one angle) , $ n_{2} \; = \; 6 $ ($ A_{y} $, 
$ D_{\ell', \ell} $, $ D_{s' s} $, $ D_{\ell',s} $, $ D_{s', \ell} $ 
and $ D_{nn} $) and $ \theta_{i} \; = \; 19^{\circ} $. Formulae 
for the calculation of $ w_{theory}^{(j)} ( \theta_{i} ) $ can 
be found in Ref. \cite{vanderVentel_PRC60_99}. 

The optimal set for a particular reaction is defined as that 
combination of effective masses for which $ \Delta $ is a minimum, 
i.e. it is that combination of effective masses which best describes 
all the spin observable data for a particular reaction at a 
particular energy. Table~\ref{table_optimal_effecmass} displays the 
optimal effective mass combinations for the various reactions in 
Table~\ref{studied_reactions}. For the second reaction in Table~\ref{studied_reactions}
($^{40}$Ca$ ( \vec{p}, \vec{p}^{\; '}) $ at
$ T_{lab} \; = \; 200$ MeV ) no optimal 
masses are listed in Table~\ref{table_optimal_effecmass}
as there were no data on complete sets of observables from
which to extract them.
For comparison Table~\ref{table_optimal_effecmass} also 
displays the effective mass values calculated in 
Ref. \cite{hillhouse_PRC49_94}. Generally one sees that, 
for both IA1 and IA2-based models, the values of the 
optimal effective masses agree to within 20\% with the 
corresponding theoretical values. In addition the optimal
effective masses do not exhibit a systematic behavior with
respect to target mass and incident energy indicating that one cannot impose a pure plane wave model on quasielastic 
scattering. Additional effects must be included in a more 
sophisticated model.

In Figs. \ref{fig_ca40_500pp19_QEP}, \ref{fig_ca40_495pn_tnt_QEP}
and \ref{fig_ca40_200pn_cheri_QEP}
we also compare IA1- and IA2-based predictions of spin observables 
based on the optimal effective masses listed in 
Table~\ref{table_optimal_effecmass}. 
The solid and dashed lines denote the IA2 and IA1 
predictions respectively. 
Deviations of the spin observables from the free mass values 
(long-dash-short-dash) serve as an indication of the importance
of effective-mass-type nuclear medium effects for quasielastic 
scattering. Generally one sees that both optimal IA1 and IA2 
predictions are very close to the free mass 
calculations indicating the insensitivity
of quasielastic spin observables to effective-mass-type medium effects. 

It is convenient to consider the spin observables in three 
different groups. Firstly, the spin observables $ D_{\ell' \ell}, 
D_{s' s}, D_{s' \ell} $ and $ D_{\ell' s} $.
For the whole energy range between 200 and 500 MeV both IA1 and
IA2 optimal effective masses provide an adequate description 
at the quasielastic peak. For the $ ( \vec{p}, \vec{n} ) $ observables 
the description is not as good as  for the $(\vec{p}, \vec{p}^{\; '})$
observables.

Next we focus on $D_{n n}$. The description of $ D_{nn} $ 
becomes problematic for both $ ( \vec{p}, \vec{p}^{\; '} ) $ 
and $ ( \vec{p}, \vec{n} ) $ scattering as the energy is lowered. 
For the $ ( \vec{p}, \vec{p}^{\; '} ) $ reaction the data point shifts 
away from the effective mass band as the energy is lowered, while 
for the $ ( \vec{p}, \vec{n} ) $ reaction the theoretical 
calculation exhibits an oscillatory motion at 495 MeV which 
causes it to miss the data. At 200 MeV there is still a variation 
with respect to laboratory scattering angle in the theoretical 
calculation whereas the data are quite flat. A possible explanation
for the latter discrepancy is the exclusion of distortions and
recoil effects in our model. 

Lastly, the analyzing power $ A_{y} $ is considered. 
In the IA2 representation of the NN scattering matrix 
the optimal effective mass set does not provide a good 
description of the $ A_{y} $ data at the quasielastic peak for 
the reaction $ ( \vec{p}, \vec{p}^{\; '} ) $ at 500 MeV. 
(It may even be better described by some other specially chosen, 
but realistic pair of effective masses.) Furthermore, as the 
energy is lowered, the $ A_{y} $ data point shifts away from 
the effective mass band. The $ ( \vec{p}, \vec{n} ) $ data 
for $ A_{y} $ are, however, much better described by the 
optimal IA2 set.

The failure of the IA2-based model to predict $A_{y}$ and
$D_{n n}$ for $ ( \vec{p}, \vec{p}^{\; '} ) $ scattering
at 200 MeV calls for a more sophisticated treatment
of nuclear distortions and recoil effects. To this end we 
have developed a relativistic distorted wave model
for quasielastic scattering \cite{hillhouse_PhD_99}; Numerical results will be presented in a future paper. 
Furthermore, since that distortions
play a more prominent role at low energies, the measurement
of a complete set of $ ( \vec{p}, \vec{p}^{\; '} ) $
spin observables at 200 MeV will be extremely useful for
checking the validity of our distorted model. The latter
measurements will also complement the existing
$( \vec{p}, \vec{n})$ data measured at the Indiana 
University Cyclotron Facility \cite{hautala_PhD_98}.

Calculations have been performed for all the reactions listed in Table~\ref{studied_reactions} as a function of energy transferred to the nucleus: these results are available from the authors on request. Conclusions based on the latter are consistent with the present investigation at the centroid of the quasielastic peak.

\section{Conclusion}
\label{section_conclusions}
In this investigation effective projectile and target nucleon 
masses were treated as free parameters and it was found that 
{\em no effective mass combination could describe both 
$ ( \vec{p}, \vec{p}^{\; '} ) $ and $ ( \vec{p}, \vec{n} ) $ 
scattering observables.} Even though the IA2 treatment of 
medium effects (within the RPWIA framework) is the most advanced 
to date, it still fails to describe all observables; the glaring 
example being the prediction of $ A_{y} $ for 
$ ( \vec{p}, \vec{p}^{\; '} ) $ scattering as the energy 
is lowered from 500 {\rm MeV} to 200 {\rm MeV}. 
In general it is seen that IA2-based effective-mass predictions
are close to the corresponding free values, whereas the ambiguous
IA1 representation severely overestimates the importance of 
effective-mass-type medium effects. Despite the 
successes of the Walecka model effective mass concept within the relativistic 
plane wave impulse approximation, the theoretical work should now start 
to include additional effects like multiple scattering, recoil effects and 
distortions of the projectile. A relativistic distorted wave model (initially employing the IA1 representation of the NN scattering 
matrix) has been presented in 
Ref. \cite{hillhouse_PhD_99}, but still needs to 
be implemented numerically. 

\acknowledgements
The authors wish to thank Professor S.J. Wallace (University of Maryland, USA) for providing the IA2 invariant amplitudes used in the present calculations.
The financial assistance to B.I.S.v.d.V by the Harry Crossley 
Foundation, the South African FRD and the National Accelerator Centre 
is gratefully acknowledged.

\begin{figure}
\caption{ Values of $ A_{y} $ and $ D_{i'j} $ versus $ \theta_{lab} $ for
$^{40}$Ca $ ( \vec{p}, \vec{p}^{\; '} ) $ at $ T_{lab} = 500 \, \mbox{{\rm MeV}} $. Solid and dashed lines represent 
the calculations with optimal effective mass values in respectively the IA2
and IA1 representations. The hatched bands denote the range of values which
result from varying $ \frac{M_{1}}{M} $ and $ \frac{M_{2}}{M} $ over the full
range (see text): The straight line hatch pattern denotes the IA1 model; the 
dotted hatch pattern the IA2 model. The long-dash--short-dash lines represent 
the free mass values. Data (at $ \theta_{lab} = 19^{\circ} $) 
are from Ref. [9].
\label{fig_ca40_500pp19_QEP}}
\end{figure}

\begin{figure}
\caption{For this reaction, $^{40}$Ca $ ( \vec{p}, \vec{p}^{\; '} ) $ at $ T_{lab} \, = \, 200 $ MeV and $ \theta_{lab} \, = \, 30^{\circ} $ only a free mass calculation (denoted by the solid line) was performed due the lack of a complete set of spin observables. The data are form Ref. [10].
\label{fig_ca40_200pp30_QEP}}
\end{figure}

\begin{figure}
\caption{Same as Fig. 1 but for the reaction $^{40}$Ca $ ( \vec{p}, \vec{n} ) $ at $ T_{lab} \, = \, 495 $ MeV and $ \theta_{lab} \, = \, 18^{\circ} $ and $ 27^{\circ} $. The data are from Ref. [11].
\label{fig_ca40_495pn_tnt_QEP}}
\end{figure}

\begin{figure}
\caption{ Same as Fig. 1 but for the reaction $^{40}$Ca $ ( \vec{p}, \vec{n} ) $ at $ T_{lab} \, = \, 200 $ MeV and $ \theta_{lab} \, = \, 24^{\circ} $, $ 37^{\circ} $ and $ 48^{\circ} $. The data are from Ref. [12].
\label{fig_ca40_200pn_cheri_QEP}}
\end{figure}

\begin{table}
\caption{Experimental data for which calculations were done at the quasielastic peak (as a function of laboratory scattering angle) and as a function of energy transfer}
\label{studied_reactions}
\begin{tabular}{|c|l|l|l|}
\ \ \ \ \ \ Reaction\ \ \ \ \ \ 
& $ T_{lab} $ (MeV)\hspace{0.8cm} & $ \theta_{lab} $ (degrees)\hspace{0.8cm}
 & Reference \hspace{0.8cm}
\\
[2mm] \hline
 $^{40}$Ca$ ( \vec{p}, \vec{p}^{\; '} ) $ & 500 & 19 & \cite{carey_PRL53_84}
\\
[2mm] \hline
$^{40}$Ca$ ( \vec{p}, \vec{p}^{\; '} ) $ & 200 & 30 & \cite{carman_PhD_95}
\\
[2mm] \hline
$^{40}$Ca$ ( \vec{p}, \vec{n}) $ & 495 & 18, 27 & \cite{taddeucci98}
\\
[2mm] \hline
$^{40}$Ca$ ( \vec{p}, \vec{n}) $ & 200 & 24, 37, 48 & \cite{hautala_PhD_98}
\\
[2mm] \hline
$^{12}$C$ ( \vec{p}, \vec{p}^{\; '} ) $ & 420 & 24 & \cite{chan_NPA510_90}
\\
[2mm] \hline
$^{12}$C$ ( \vec{p}, \vec{p}^{\; '} ) $ & 290 & 30 & \cite{chan_NPA510_90}
\\
[2mm] \hline
$^{54}$Fe$ ( \vec{p}, \vec{p}^{\; '} ) $ & 290 & 20 & \cite{hausser_PRL61_88}
\\
[2mm] \hline
$^{208}$Pb$ ( \vec{p}, \vec{n}) $ & 200 & 24, 37, 48 & \cite{hautala_PhD_98}
\\
\end{tabular}
\end{table}

\begin{table}
\caption{Values of optimal effective mass combinations, $ \displaystyle{( \frac{M_{1}}{M}, \frac{M_{2}}{M} )} $, extracted at the quasielastic peak. The last column refers to the effective mass combinations which are calculated theoretically [18].}
\label{table_optimal_effecmass}
\begin{tabular}
{|c|c|c|c|c|}
Reaction & $T_{lab}$ (MeV)
& IA1 & IA2 
& Theory \\
\hline
 &  & $\frac{M_1}{M}$\hspace{2.3cm}$\frac{M_2}{M}$\ \ \ \ &
$\frac{M_1}{M}$\hspace{2.3cm} $\frac{M_2}{M}$\ \ \ \ &
$\frac{M_1}{M}$\hspace{2.3cm} $\frac{M_2}{M}$\ \ \ \ \\
\hline
$^{40}$Ca$ ( \vec{p}, \vec{p}^{\; '} ) $ & 500 & 
0.96 \hspace{2.0cm} 0.96 \ \ \ \ & 1.0 \hspace{2.0cm} 0.86\ \ \ \
& 0.89 \hspace{2.0cm} 0.82 
\\
[2mm] \hline
$^{40}$Ca$ ( \vec{p}, \vec{n} ) $ & 495 & 
0.75 \hspace{2.0cm} 0.75\ \ \ \ & 0.85 \hspace{2.0cm} 0.85 \ \ \ \
& 0.89 \hspace{2.0cm} 0.82
\\
[2mm] \hline
$^{40}$Ca$ ( \vec{p}, \vec{n} ) $ & 200 & 
1.0 \hspace{2.0cm} 0.87\ \ \ \ & 0.93 \hspace{2.0cm} 0.92 \ \ \ \
& 0.83 \hspace{2.0cm} 0.75
\\
[2mm] \hline
$^{12}$C$ ( \vec{p}, \vec{p}^{\; '} ) $ & 
420 & 0.88 \hspace{2.0cm} 0.88 \ \ \ \ & 0.88 \hspace{2.0cm} 0.76 \ \ \ \
& 0.86 \hspace{2.0cm} 0.79
\\
[2mm] \hline
$^{12}$C$ ( \vec{p}, \vec{p}^{\; '} ) $ & 
290 & 0.93 \hspace{2.0cm} 0.93\ \ \ \ & 1.0 \hspace{2.0cm} 0.92 \ \ \ \
& 0.83 \hspace{2.0cm} 0.77
\\
[2mm] \hline
$^{54}$Fe$ ( \vec{p}, \vec{p}^{\; '} ) $ & 
290 & 0.87 \hspace{2.0cm} 0.87 \ \ \ \ & 0.86 \hspace{2.0cm} 0.74 \ \ \ \
& 0.83 \hspace{2.0cm} 0.77
\\
[2mm] \hline
$^{208}$Pb$ ( \vec{p}, \vec{n}) $ & 
200 & 0.94 \hspace{2.0cm} 0.84 \ \ \ \ & 0.94 \hspace{2.0cm} 0.94 \ \ \ \
 & 0.85 \hspace{2.0cm} 0.83
\\
\end{tabular}
\end{table}

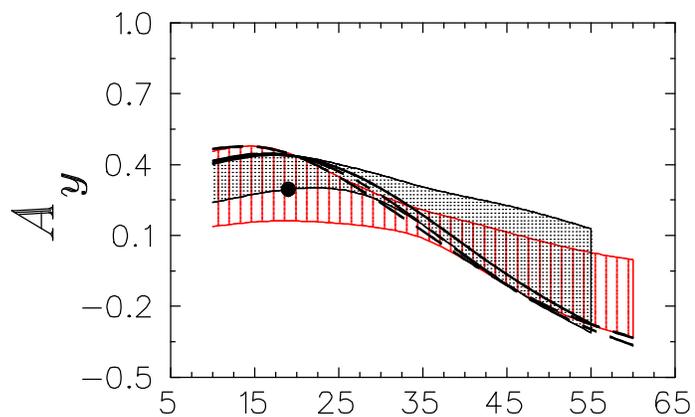
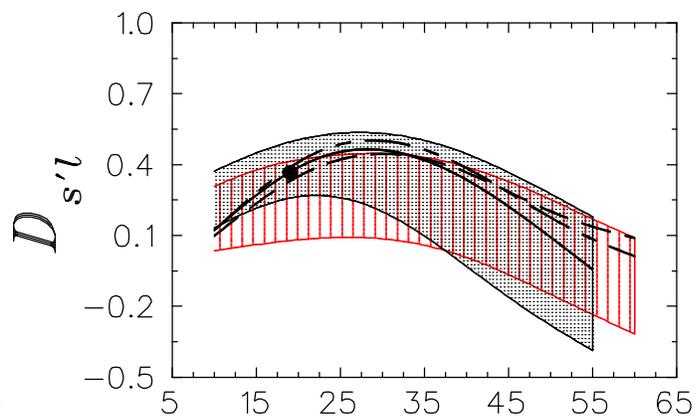
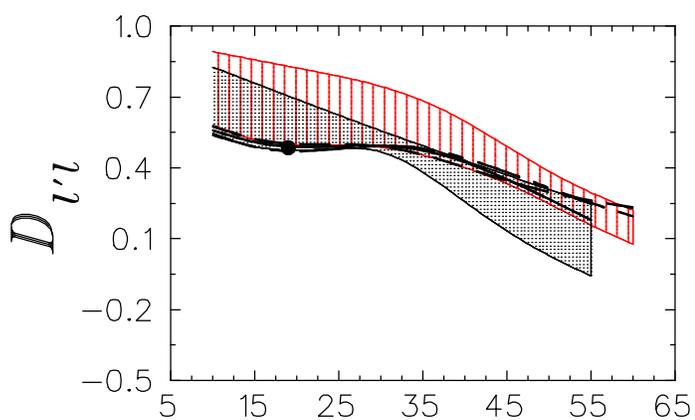
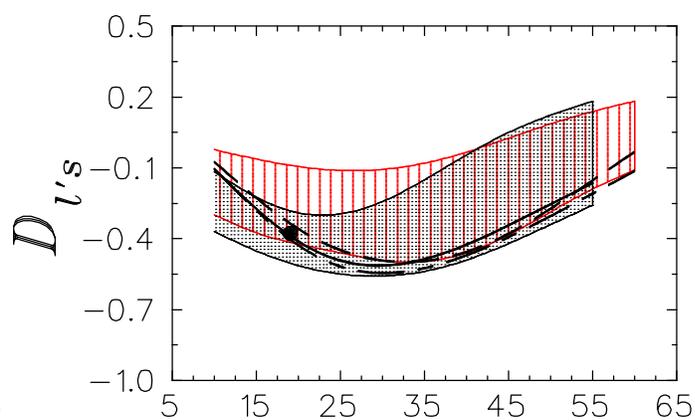
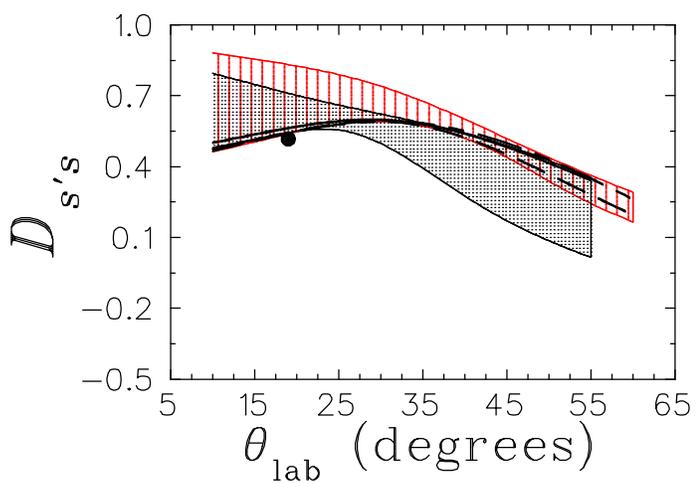
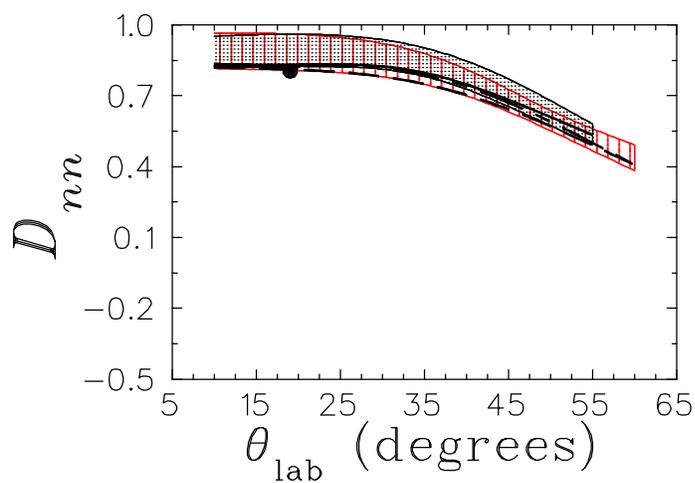
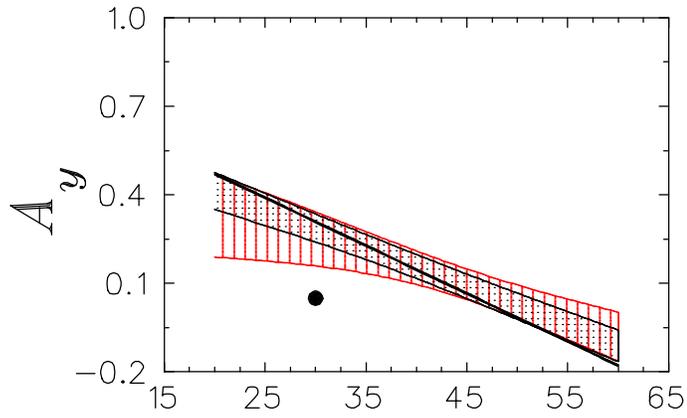
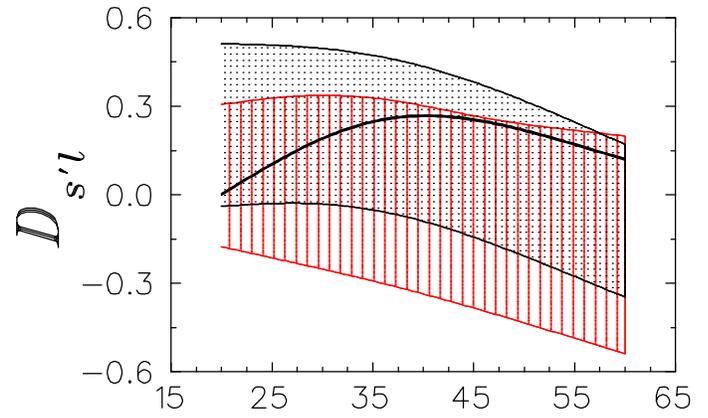
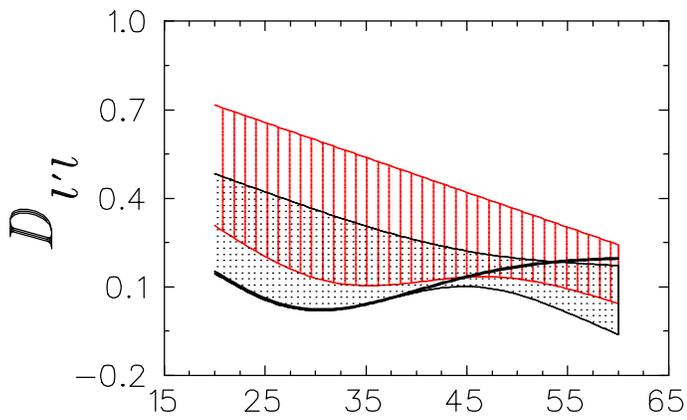
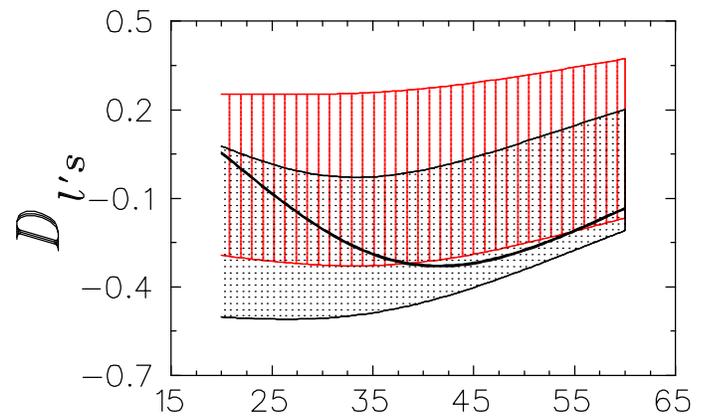
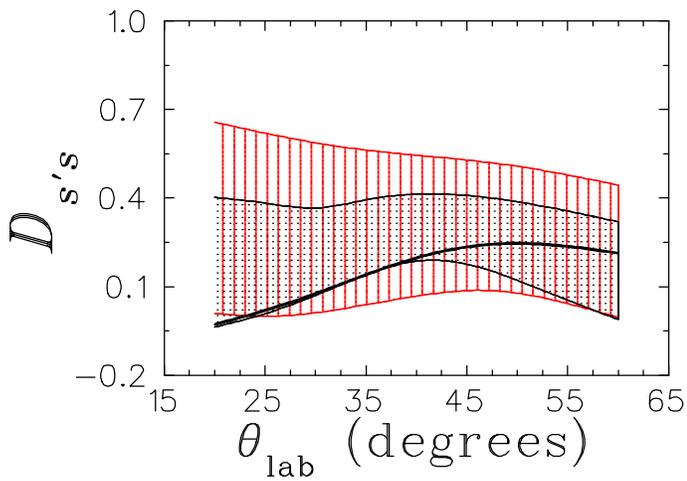
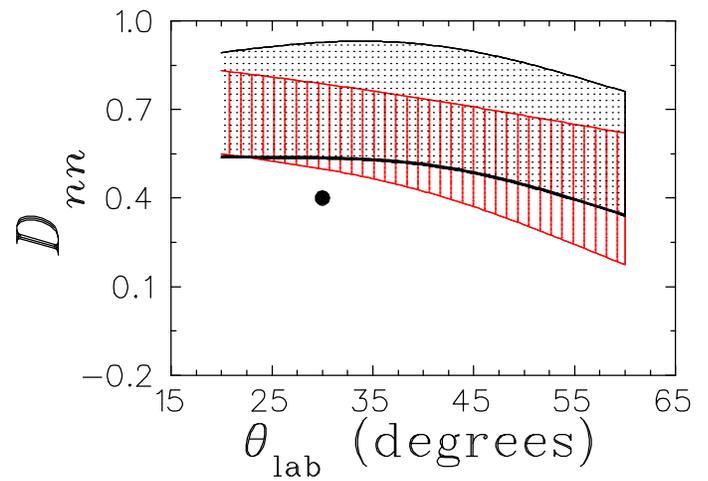
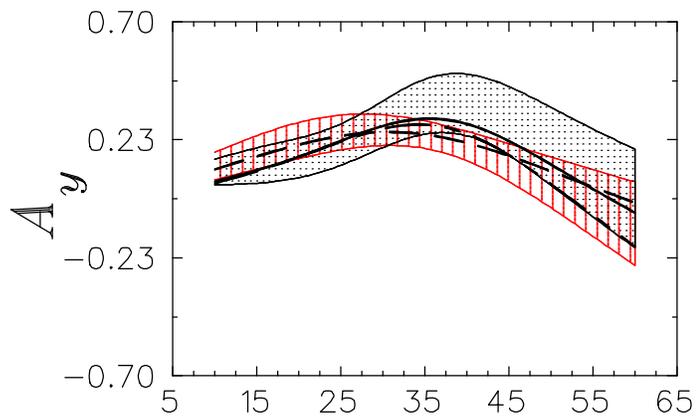
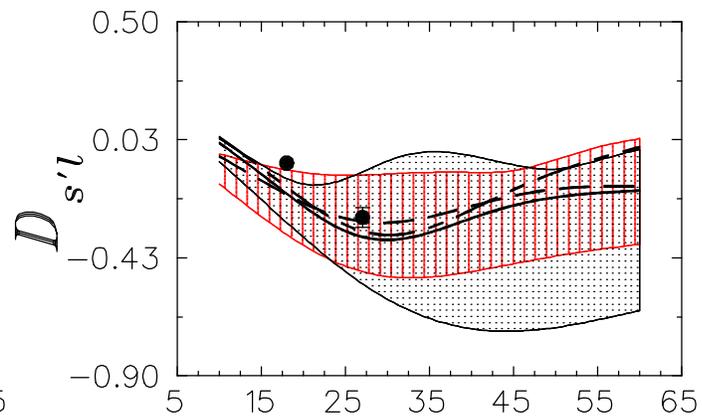
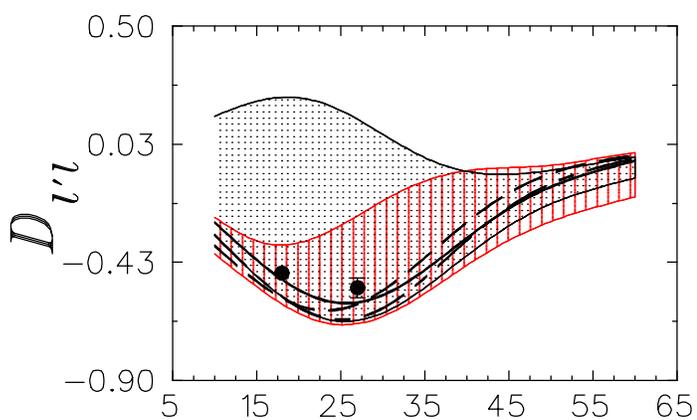
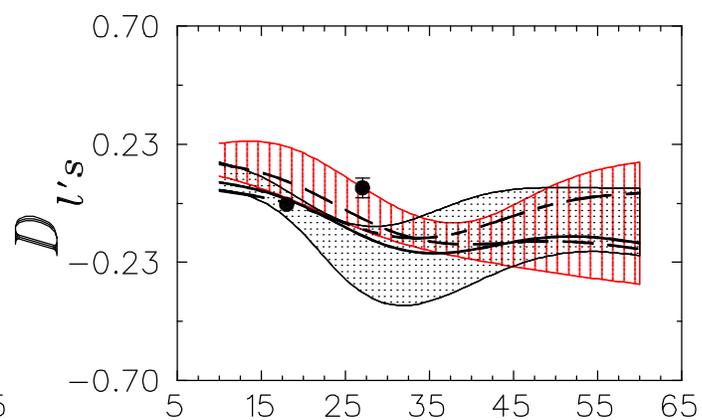
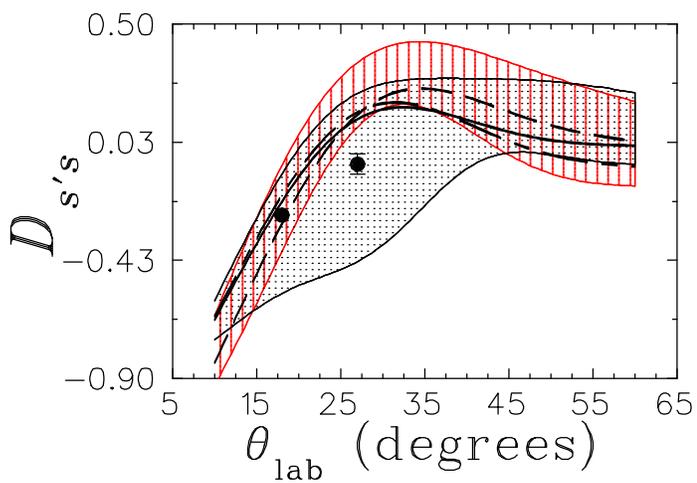
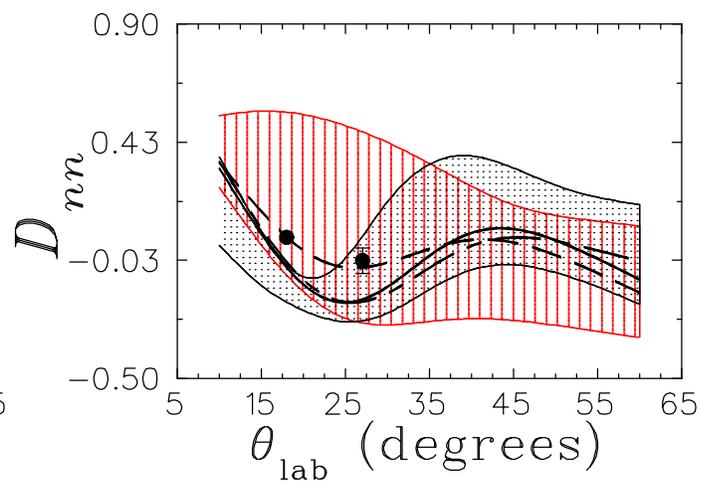
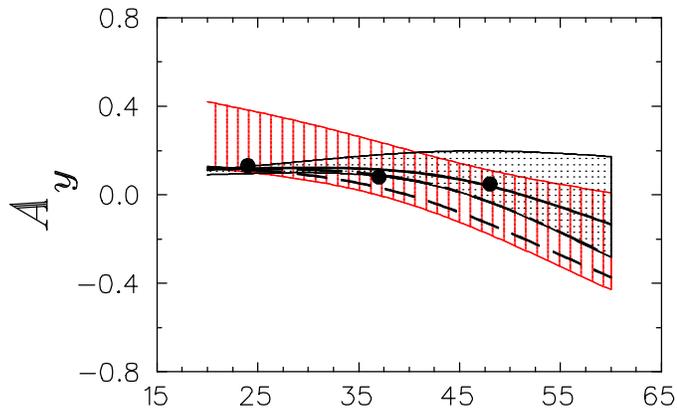
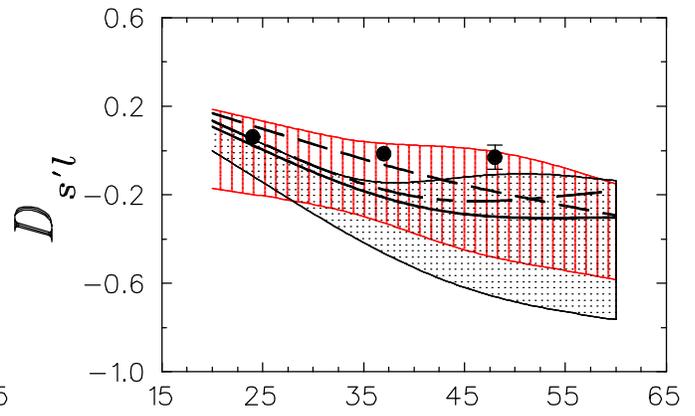
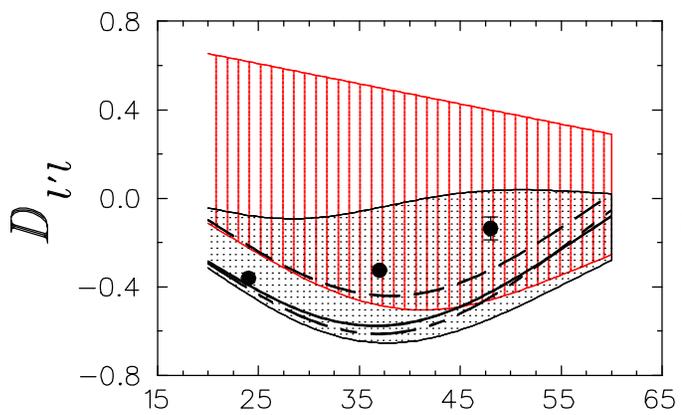
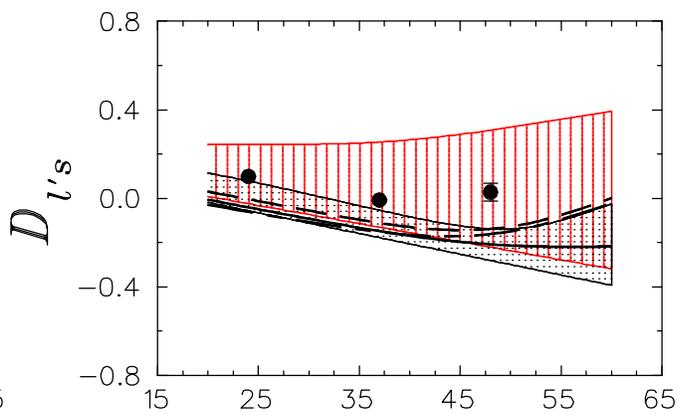
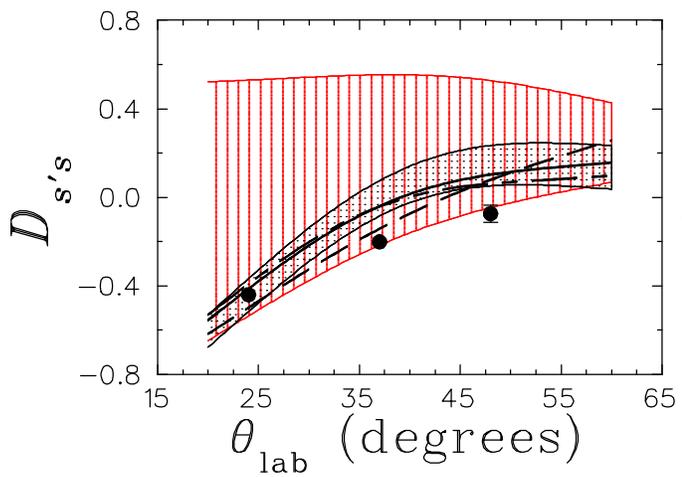
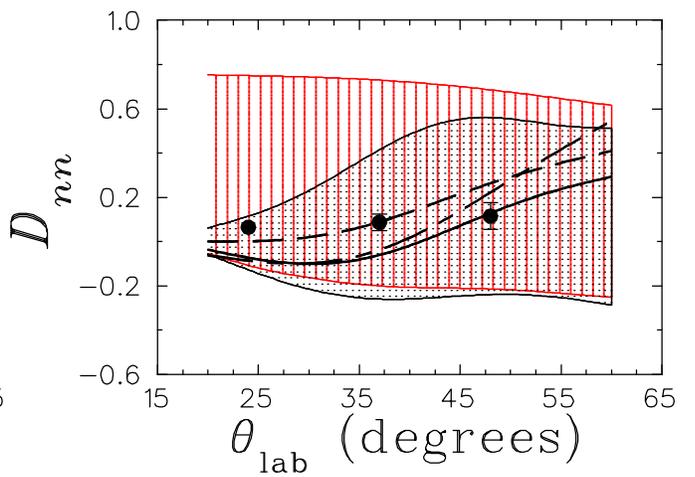
\end{document}